\documentclass[aps,prl,twocolumn,showpacs,superscriptaddress]{revtex4}
\usepackage{graphicx}

\begin{document}

\title{Effect of electron-doping on spin excitations of underdoped BaFe$_{1.96}$Ni$_{0.04}$As$_{2}$}
\author{Leland W. Harriger}
\affiliation{
Department of Physics and Astronomy, The University of Tennessee, Knoxville, Tennessee 37996-1200, USA
}
\author{Astrid Schneidewind}
\affiliation{Technische Universit\"{a}t Dresden,
 Institut f\"{u}r Festk\"{o}rperphysik, 01062 Dresden, Germany
}
\author{Shiliang Li}
\affiliation{
Department of Physics and Astronomy, The University of Tennessee, Knoxville, Tennessee 37996-1200, USA
}
\affiliation{
Institute of Physics, Chinese Academy of Sciences, Beijing 100080, China
}
\author{Jun Zhao}
\affiliation{
Department of Physics and Astronomy, The University of Tennessee, Knoxville, Tennessee 37996-1200, USA
}
\author{Zhengcai Li}
\affiliation{
Institute of Physics, Chinese Academy of Sciences, Beijing 100080, China
}

\author{Wei Lu}
\affiliation{
Institute of Physics, Chinese Academy of Sciences, Beijing 100080, China
}
\author{Xiaoli Dong}
\affiliation{
Institute of Physics, Chinese Academy of Sciences, Beijing 100080, China
}

\author{Fang Zhou}
\affiliation{
Institute of Physics, Chinese Academy of Sciences, Beijing 100080, China
}
\author{Zhongxian Zhao}
\affiliation{
Institute of Physics, Chinese Academy of Sciences, Beijing 100080, China
}
\author{Jiangping Hu}
\affiliation{Department of Physics, Purdue University, West Lafayette, IN 47907}
\author{Pengcheng Dai}
\email{daip@ornl.gov}
\affiliation{
Department of Physics and Astronomy, The University of Tennessee, Knoxville, Tennessee 37996-1200, USA
}
\affiliation{
Institute of Physics, Chinese Academy of Sciences, Beijing 100080, China
}
\affiliation{Neutron Scattering Science Division, Oak Ridge National Laboratory, Oak Ridge, Tennessee 37831-6393, USA}

\begin{abstract}
We use neutron scattering to study magnetic order and spin excitations in BaFe$_{1.96}$Ni$_{0.04}$As$_{2}$.
On cooling, the system first changes the lattice symmetry from tetragonal to orthoromhbic near $\sim$97 K,
and then orders antiferromagnetically at $T_N=91$ K before developing weak superconductivity below $\sim$15 K.
Although superconductivity appears to
co-exist with static antiferromagnetic order from transport and neutron diffraction measurement,
inelastic neutron scattering experiments reveal that magnetic excitations do not respond to superconductivity. Instead, the effect of electron-doping is to reduce the $c$-axis exchange coupling in BaFe$_2$As$_2$ and
induce quasi two-dimensional spin excitations.  These results suggest that transition from three-dimensional spin waves to two-dimensional spin excitations by electron-doping is important for the
separated structural/magnetic phase transitions and high-temperature superconductivity in iron arsenides.
\end{abstract}


\maketitle

Antiferromagnetism is relevant to high temperature (high-$T_c$) superconductivity in copper oxides and iron arsenides because superconductivity arises from
electron- or hole-doping of their static antiferromagnetic (AF) ordered parent compounds \cite{palee,birgeneau,eschrig,kamihara,rotter,sefat,ljli,cruz,jzhao1}.
In the case of cuprates, spin waves of the parent materials can be very well described by a local moment Heisenberg Hamiltonian and spin excitations in optimally doped superconductors are dominated by a neutron spin resonance centered at the AF ordering wavevector \cite{palee,birgeneau,eschrig}. For undoped iron arsenides such as $A$Fe$_2$As$_2$ ($A=$Ba, Sr, Ca) with a spin structure of Fig. 1a \cite{jeff},
spin waves consist of a large anisotropy gap at the AF zone center [$\Delta(1,0,1)\leq 9.8$ meV]
and excitations extend up to $\sim$200 meV \cite{jzhao2,ewings,rob,matan,jzhao3}.
For optimally doped superconductors \cite{rotter,sefat,ljli}, the gapped spin wave excitations were replaced by a gapless continuum of scattering in the normal state and a neutron spin resonance below $T_c$ \cite{christianson,lumsden,chi,slli}. Since spin fluctuations may play a crucial role in the electron pairing and superconductivity of iron arsenides \cite{mazin,si,hu,xu,hu2,wang,weng,chubukov}, it is imperative to determine how the spin dynamics of the undoped AF parent compounds evolve as they are tuned toward optimally doped superconductivity by electron or hole doping.

In the undoped state, BaFe$_{2}$As$_{2}$
exhibits simultaneous structural and magnetic phase transitions below $T_s=T_N=143$ K,
changing the crystal lattice symmetry from the high-temperature tetragonal to low-temperature orthorhombic phase  \cite{jeff}. Upon Co-doping to induce electrons onto the FeAs plane, the combined AF and structural phase transitions were split into two distinct transitions and the electronic phase diagram in the lower Co-doping region displays coexisting static AF order with the superconductivity \cite{ni,chu}.  Although recent neutron scattering experiments confirmed that the upper transition is structural
and the AF order occurs at lower temperature \cite{pratt,lester,christianson2}, its microscopic origin is still unknown. More importantly, it is unclear what happens to the spin waves of BaFe$_2$As$_2$ when electrons are doped into these materials.  While Pratt {\it et al.} \cite{pratt} reported gapless normal state spin excitations
for BaFe$_{1.906}$Co$_{0.094}$As$_2$, measurements on BaFe$_{1.92}$Co$_{0.08}$As$_2$ suggest gapped normal state excitations \cite{christianson2}. On cooling below $T_c$, both materials reveal a reduction in the static ordered AF moment and the appearance of a spin resonance \cite{pratt,christianson2}.

\begin{figure}[t]
\includegraphics[scale=.4]{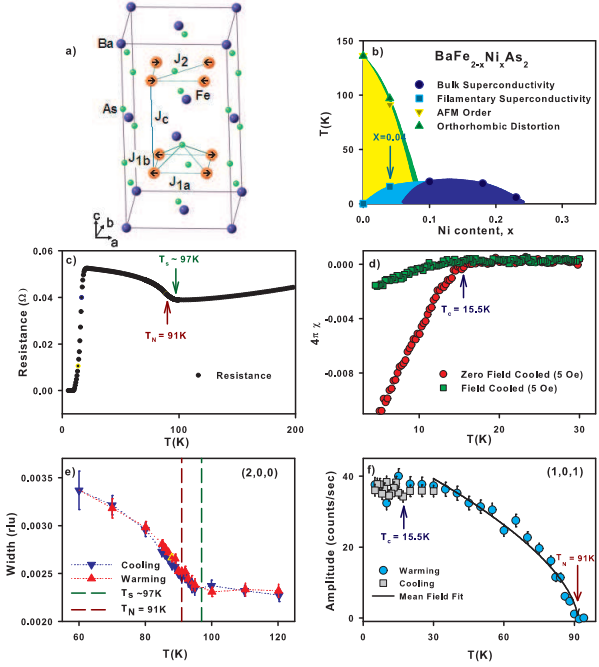}
\caption{
(color online). (a) Diagram of the parent compound BaFe$_{2}$As$_{2}$ with Fe spin ordering and magnetic exchange couplings depicted. We use the same unit cell for BaFe$_{1.96}$Ni$_{0.04}$As$_{2}$.  (b) Electronic phase diagram from Ref. \cite{ljli}.  (c)
Temperature dependence of the resistance showing anomalies at $T_s$, $T_N$, and $T_c$. (d) Temperature
dependence of the Meissner and shielding signals on a small crystal (field cooled $4\pi\chi=-0.001$ at 4.5 K).
(e) Temperature dependence of the structural distortion of the lattice as determined by tracking the
width of the $(2,0,0)$ nuclear Bragg peak using $\lambda/2$ scattering without Be filter. (f) Magnetic order parameter determined by $Q$-scans around $(1,0,1)$ magnetic Bragg peak above background. The solid line shows
order parameter fit using $\phi^2\propto(1-T/T_N)^{2\beta}$ with $T_N=91.3\pm0.7$ K and $\beta=0.3\pm0.02$.
 }
\end{figure}

To compare with the results obtained on Co-doped BaFe$_{2}$As$_{2}$ \cite{ni,chu,pratt,lester,christianson2}, we carried out neutron scattering experiments on Ni-doped  BaFe$_{1.96}$Ni$_{0.04}$As$_2$ ($T_c\approx 15$ K, Figs. 1c,1d) \cite{ljli}. In contrast to the results on
Co-doped materials \cite{pratt,christianson2},  we find that the static AF order and spin excitations in BaFe$_{1.96}$Ni$_{0.04}$As$_2$ do not respond to the occurence  of superconductivity. Instead, the effect of electron-doping is to significantly reduce the $c$-axis exchange coupling and change the three-dimensional (3D)
spin waves of BaFe$_{2}$As$_{2}$ into quasi two-dimensional (2D).  These results suggest that
the separated structural/magnetic phase transition and the appearance of bulk superconductivity upon doping may be associated with the diminishing spin anisotropy gap and the 3D to 2D transition of the spin excitations.

\begin{figure}[t]
\includegraphics[scale=.4]{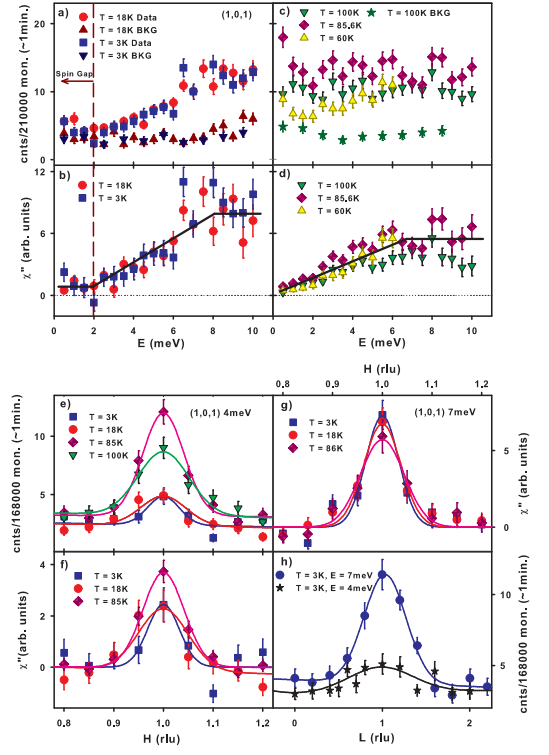}
\caption{
(color online). (a) Energy scans at $Q = (1,0,1)$ and $Q=(1.2,0,1)$ above and below $T_{c}$.
(b) $\chi^{\prime\prime}(Q,\omega)$ at $Q = (1,0,1)$.
(c) Energy scans at higher temperatures and, (d) the corresponding $\chi^{\prime\prime}(Q,\omega)$.
(e) $Q$-scans along the $[H,0,1]$ direction at 4 meV and different temperatures.  Fourier transforms of the
Gaussian peaks give minimum correlation lengths of $\xi=57\pm4$ \AA\.  (f) Estimated $\chi^{\prime\prime}(Q,\omega)$ at 4 meV.
(g) $\chi^{\prime\prime}(Q,\omega)$ at 7 meV with minimum correlation length of $\xi=54\pm6$ \AA.
(h) Low temperature $Q$-scans along the $[1,0,L]$ direction ($c$-axis) at 4 meV and 7 meV.
The minimum
dynamic spin correlation lengths are $\xi\approx 14\pm5$ and $21\pm2$ \AA\
for 4 and 7~meV, respectively.
 }
\end{figure}

Using the self-flux method \cite{ljli}, we grew a $\sim1$ gram
single crystal of BaFe$_{1.96}$Ni$_{0.04}$As$_{2}$ with
 an in-plane and out-of-plane mosaic of 1.74$^{\circ}$ and 2.20$^{\circ}$ full-width at half maximum (FWHM), respectively. We defined the wave vector $Q$ at ($q_x$, $q_y$, $q_z$) as $(H,K,L) = (q_xa/2\pi, q_yb/2\pi, q_zc/2\pi)$ reciprocal lattice units (rlu) using the orthorhombic magnetic unit cell
(space group Fmmm), where $a = 5.5$ \AA, $b = 5.4$ \AA, and $c = 12.77$ \AA.
 We performed our neutron scattering experiment on the PANDA cold triple-axis spectrometer at the Forschungsneutronenquelle Heinz Maier-Leibnitz (FRM II), TU Munchen, Germany as described earlier \cite{chi}.
Our sample was aligned in the $[H,0,L]$ zone inside a closed cycle refrigerator.

\begin{figure}[t]
\includegraphics[scale=.4]{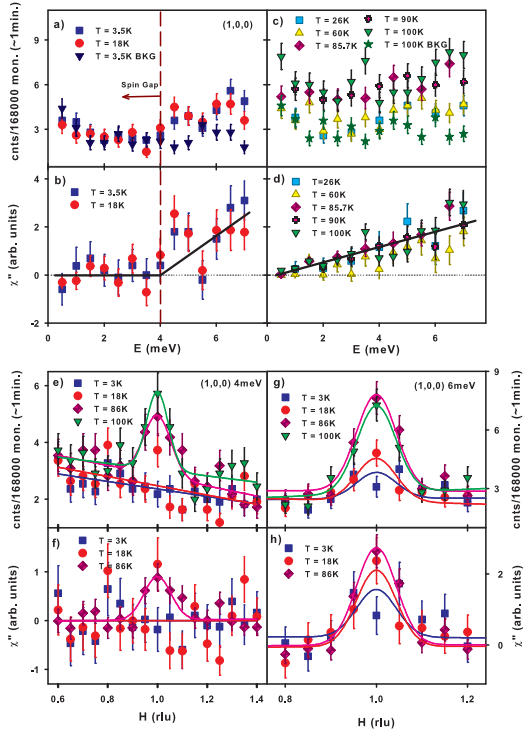}
\caption{(color online).
(a) Energy scans at $Q=(1,0,0)$ and $Q=(1.4,0,0)$ from 0.5 meV to 7 meV at 3.5 K and 18 K.
(b) Background corrected $\chi^{\prime\prime}(Q,\omega)$ showing clear evidence for a 4 meV spin gap, larger than
that at $Q=(1,0,0)$.
(c) Temperature dependence of the signal [$Q=(1,0,0)$] and background [$Q=(1.4,0,0)$] scattering at various temperatures. (d) $\chi^{\prime\prime}(Q,\omega)$ at 4 meV.
(e) $Q$-scans along the $[H,0,0]$ direction at 4 meV and different temperatures.  A peak centered at $Q=(1,0,0)$ appears above 80 K. (f) Background corrected $\chi^{\prime\prime}(Q,\omega)$.
(g) Temperature dependence of the $Q$-scans along the $[H,0,0]$ direction at 6 meV.  The scattering has a peak
at 3 K. (h) Temperature dependence of the $\chi^{\prime\prime}(Q,\omega)$
at 6 meV.
 }
\end{figure}

Figures 1c and 1d show the resistivity and susceptibility data. The resistivity shows clear anomalies near 97 K and 91 K before superconductivity sets in below $\sim$15 K (Fig. 1c). Although the presence of superconductivity below $T_c\approx15$ K is confirmed in the susceptibility measurement (Fig. 1d), the weak Meissner effect
suggests superconducting volume fraction of less than 0.2\%. Similar to Co-doped BaFe$_2$As$_2$ \cite{pratt,lester,christianson2}, we find
that the tetragonal to orthorhombic structural transition happens at 97 K while the AF order occurs below $T_N=91$ K (Figs. 1e and 1f). However, in contrast with Co-doped materials \cite{pratt,christianson2},  superconductivity has no influence on the static AF order of BaFe$_{1.96}$Ni$_{0.04}$As$_{2}$ (Fig. 1f). This is consistent with the fact that our sample has a lower electron-doping to the FeAs-plane than those of Refs. \cite{pratt,christianson2}.

\begin{figure}[t]
\includegraphics[scale=.3]{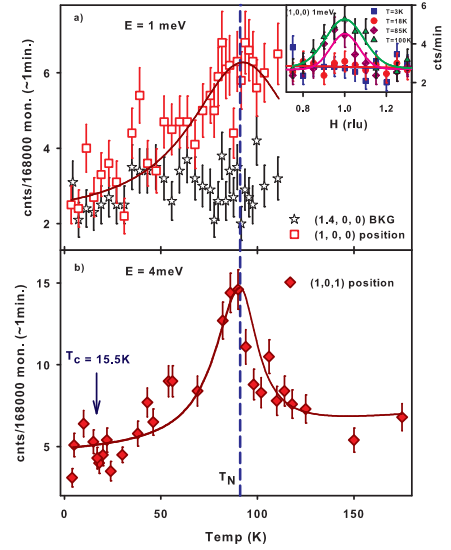}
\caption{(color online).
(a) Temperature dependence of the 1 meV scattering at the signal
$Q=(1,0,0)$ and background $Q=(1.4,0,0)$ positions.  The inset shows $Q$-scans along
the $[H,0,0]$ at 1 meV and different temperatures.  The scattering shows
no anomaly across $T_c$ but clearly peak at $T_N$. (b)
Temperature dependence of the scattering at 4 meV and $Q=(1,0,1)$ again peaks at $T_N$.
 }
\end{figure}

In the undoped BaFe$_2$As$_2$, spin waves have an anisotropy gap 
about 8 meV at $Q=(1,0,1)$ [$\Delta(1,0,1)=8$ meV] \cite{ewings,matan}.
For optimally Co and Ni doped materials, spin excitations are gapless in the normal state \cite{lumsden,chi} and  superconductivity-induced spin gaps open below $T_c$ \cite{slli}.
Figure 2a shows the
constant-$Q$ scans at the $Q=(1,0,1)$ (signal) and
$Q=(1.2,0,1)$ (background) positions above and below $T_c$ for BaFe$_{1.96}$Ni$_{0.04}$As$_{2}$.  Figure 2b plots the imaginary part of the dynamic susceptibility $\chi^{\prime\prime}(Q,\omega)$ after correcting for background and Bose population factor. We find that $\chi^{\prime\prime}(Q,\omega)$ has
a 2 meV spin gap and is not affected by superconductivity. Figures 2c and 2d 
reveal that the magnetic intensity increase with increasing temperature below $T_N$
is due mostly to the Bose population factor.
These results are confirmed by $Q$-scans along the $[H,0,1]$ direction at different temperatures (Figs. 2e-2g),
which display well-defined peaks at $Q=(1,0,1)$ that have similar widths to the undoped BaFe$_2$As$_2$ at 10 meV \cite{matan}. Figure 4h shows $Q$-scans along the $c$-axis $[1,0,L]$ direction.
In contrast to the in-plane spin excitations, the $c$-axis
 spin-spin correlations are much broader for BaFe$_{1.96}$Ni$_{0.04}$As$_{2}$, thus suggesting
2D nature of the excitations.

Further evidences for 2D spin excitations in BaFe$_{1.96}$Ni$_{0.04}$As$_{2}$ are summarized in Fig. 3.
Assuming spin excitations in BaFe$_{2-x}$Ni$_{x}$As$_{2}$ can be described by an effective Heisenberg Hamiltonian,
the spin anisotropy gaps at $Q=(1,0,1)$ and $Q=(1,0,0)$ are $\Delta(1,0,1)=2S[(J_{1a}+2J_2+J_c+J_s)^2-(J_c+J_{1a}+2J_2)^2]^{1/2}$ and $\Delta(1,0,0)=2S[(2J_{1a}+4J_2+J_s)(2J_c+J_s)]^{1/2}$, respectively \cite{jzhao2,ewings,matan,jzhao3}.  Here
$S$ is the magnetic spin ($=1$); $J_{1a}$, $J_2$, $J_c$, $J_s$ are effective in-plane nearest-neighbor, next nearest-neighbor, $c$-axis, and magnetic single ion anisotropy couplings, respectively (Fig. 1a). For BaFe$_2$As$_2$, we estimate $\Delta(1,0,1)=7.8$ meV and $\Delta(1,0,0)=20.2$ meV assuming $J_{1a}=36$, $J_2=18$, $J_c=0.3$, $J_s=0.106$ meV \cite{ewings,matan,jzhao3}.
Upon electron doping to form BaFe$_{1.96}$Ni$_{0.04}$As$_{2}$, these spin gap values have been reduced to $\Delta(1,0,1)=2$ meV and $\Delta(1,0,0)=4$ meV (Figs. 2b and 3b).
Since such electron-doping hardly changes the in-plane $Q$-scan widths compared to that of the
undoped BaFe$_2$As$_2$ (Figs. 2e-g, 3e,3g) \cite{ewings,matan}, it should only
slightly modify the in-plane exchange couplings. Assuming that $J_{1a}$ and $J_2$ are unchanged in BaFe$_{1.96}$Ni$_{0.04}$As$_{2}$, the observed $\Delta(1,0,1)=2$ meV and $\Delta(1,0,0)=4$ meV would
correspond to $J_c=0.01$ meV and $J_s=0.007$ meV, suggesting a rapid suppression
of $c$-axis exchange coupling and magnetic single ion anisotropy with electron doping.

In Ref. \cite{christianson2}, it was argued that spin anisotropy for BaFe$_{1.92}$Co$_{0.08}$As$_{2}$ is similar to that of the BaFe$_2$As$_2$, meaning that
the reduction in spin gap at $Q=(1,0,1)$ arises mostly from reduced $J_{1a}$ and $J_{2}$.
Assuming the best fitted values of $S(J_{1a}+2J_2)=32$ meV and $SJ_c=0.34$ meV \cite{christianson2},
we expect $\Delta(1,0,1)=5.5$ meV and $\Delta(1,0,0)=14.2$ meV with $SJ_s=0.106$ meV.  These values are 
clearly different from the observation.  Even if we assume all exchange couplings to reduce by 50\% upon electron-doping with
$S(J_{1a}+2J_2)=32$ meV, $SJ_c=0.15$ meV, and $SJ_s=0.05$ meV, we still find $\Delta(1,0,1)=3.8$ and $\Delta(1,0,0)=10$ meV.  This suggests that the large reduction in the
$\Delta(1,0,0)$ gap values upon electron
doping is due to the reduced $J_c$ and three-dimensionality of the system.

To determine the temperature dependence of $\Delta(1,0,0)$, we show in Fig. 3c
the observed scattering at the signal $Q=(1,0,0)$ and background $(1.4,0,0)$ positions
at several temperatures. Figure 3d plots the estimated $\chi^{\prime\prime}(Q,\omega)$.
Comparing Fig. 3d with Fig. 3b,
the 4 meV spin gap $\chi^{\prime\prime}(Q,\omega)$ at 18 K vanishes upon warming to above 60 K.
These results are confirmed by $Q$-scans at 4 meV along the
$[H,0,0]$ direction (Fig. 3e).  While scans at 2 K and 18 K are featureless, the scattering at 86 K and 100 K shows clear peaks centered at $Q=(1,0,0)$.
For $Q$-scans at 6 meV,
the scattering shows well-defined peaks at all temperatures (Fig. 3g).  Converting these data into $\chi^{\prime\prime}(Q,\omega)$ in Fig. 3h confirms the results of Fig. 3d.

Finally, we show in Fig. 4a the temperature dependence of the 1 meV
scattering at the $Q=(1,0,0)$ (signal) and $Q=(1.4,0,0)$ (background) positions.
While the background scattering only increases slightly with increasing temperature and shows no anomaly across $T_N$, the scattering at $Q=(1,0,0)$ clearly peaks at $T_N$.  $Q$-scans along the $[H,0,0]$ direction at
1 meV confirm these results (the inset of Fig. 4a).  
Temperature dependence of the scattering at 4 meV and $Q=(1,0,1)$ show similar behavior (Fig. 4b).
These results suggest
that the disappearing $\Delta(1,0,1)$ and $\Delta(1,0,0)$ gaps near $T_N$ arise from critical scattering associated with the static AF order. 

To understand the separated structural and magnetic phase transitions for BaFe$_{1.96}$Ni$_{0.04}$As$_{2}$,
we note that in an effective $J_1$-$J_2$-$J_c$ model \cite{hu,xu}, the separation of the lattice and magnetic transition temperatures is controlled by the value of $J_c$ \cite{hu}.
There is only one transition temperature when $J_c$ is large. A finite separation between the two transition temperatures occurs when $J_c/J_2$ is reduced to the order of $10^{-3}$.
Our experimental result of ${J_c}/{J_2}\sim 0.5 \times 10^{-3}$ is consistent with this picture.
To quantitatively estimate the reduced $T_N$ due to the smaler $J_c$, 
we note that $T_N\sim J_2/ln(J_2/J_c)$ \cite{hu}.
Let $J_{\alpha}^0$  be the magnetic exchange values for the parental compounds, we can write $T_N^0/T_N = a[ln(a)+ln(b)+ln(c)]/ln(c)$, where $a=J^0_2/J_2, b=J^0_c/J_c,
c=J_2^0/J_c^0$. Using the experimental values of the exchange coupling parameters determined earlier, we obtain   $(J_2/J_2^0)=(1/a)\sim 0.87$, which is self-consistent with our suggestion that upon doping, the coupling between layers $J_c$ is dramatically reduced while the change of the in-plane magnetic exchange coupling is small.   These results provide a natural and consistent interpretation for our experimental observations.

In summary, we have shown that the most dramatic effect of electron-doping in BaFe$_2$As$_2$ is to transform the 3D anisotropic spin waves into 2D spin excitations. These results suggest that reduced dimensionality in spin excitations of iron asenides is important for the separated structural/magnetic phase transition and the occurance of high-$T_c$ superconductivity in these materials.

We thank C. Xu for helpful discussions. This work is supported by the U.S. NSF No. DMR-0756568, U.S. DOE BES No.
DE-FG02-05ER46202, and by the U.S. DOE, Division of Scientific
User Facilities. The work in IOP is supported by
the Ministry of Science and Technology of China and NSFC.
We further acknowledge support from DFG
within Sonderforschungsbereich 463 and from the PANDA project of
TU Dresden and FRM II.


\end{document}